\documentclass[conference]{IEEEtran}
\IEEEoverridecommandlockouts

\usepackage{cite}
\usepackage{amsmath,amssymb,amsfonts}
\usepackage{amsthm}
\usepackage{bm}
\usepackage{mathdots}
\usepackage{stfloats}

\usepackage{multirow}
\usepackage{array}
\usepackage{makecell}   
\usepackage{booktabs}   
\usepackage{siunitx}    
\usepackage[percent]{overpic} 
\usepackage{soul}

\usepackage{comment}
\usepackage{graphicx}
\usepackage{textcomp}

\usepackage{enumitem}
\usepackage{adjustbox}
\usepackage{mathtools}
\PassOptionsToPackage{hyphens}{url}
\usepackage[hidelinks]{hyperref}
\usepackage{tikz,pgfplots}
\usepackage{pgfplotstable}
\usetikzlibrary{pgfplots.fillbetween} 
\usetikzlibrary{positioning}
\usetikzlibrary{decorations.pathreplacing}
\usetikzlibrary {arrows.meta,bending,positioning}
\pgfdeclareradialshading{sphere}{\pgfpoint{0cm}{0cm}}%
{rgb(0cm)=(1,1,1);
rgb(0.7cm)=(0.7,0.1,0); rgb(1cm)=(0.5,0.05,0); rgb(1.05cm)=(1,1,1)}
\usepackage{grffile}
\pgfplotsset{compat=newest}
\usepgfplotslibrary{patchplots}
\usepgfplotslibrary{statistics}
\usetikzlibrary{backgrounds,plotmarks,fit,positioning,arrows,shapes,shapes.multipart,calc,arrows.meta}
\usetikzlibrary{fit,positioning,arrows,shapes,shapes.multipart,calc,arrows.meta,shapes.geometric,shapes.misc}
\usetikzlibrary{decorations.pathreplacing,angles,quotes,calligraphy}
\usetikzlibrary{automata}
\usetikzlibrary{arrows.meta, positioning, backgrounds, fit}
\usetikzlibrary{spy}

\allowdisplaybreaks

\usepackage{cuted}

\usepackage{vmr-symbols-vecbold-RanDetSame}
\usepackage{standard-macros}

\newcommand{\adjustedbar}[1]{\overline{\hspace{-0.1em}#1\hspace{-0.1em}}}

\usepackage{algorithm}
\usepackage{algpseudocode}

\usepackage[caption=false,font=normalsize,labelfont=sf,textfont=sf]{subfig}

\makeatletter
\newcommand\fs@betterruled{%
  \def\@fs@cfont{\bfseries}\let\@fs@capt\floatc@ruled
  \def\@fs@pre{\vspace*{8pt}\hrule height.8pt depth0pt \kern2pt}%
  \def\@fs@post{\kern2pt\hrule\relax}%
  \def\@fs@mid{\kern2pt\hrule\kern2pt}%
  \let\@fs@iftopcapt\iftrue}
\floatstyle{betterruled}
\restylefloat{algorithm}
\makeatother

\def\BibTeX{{\rm B\kern-.05em{\sc i\kern-.025em b}\kern-.08em
    T\kern-.1667em\lower.7ex\hbox{E}\kern-.125emX}}

\title{
Type-Based Unsourced Federated Learning \\With Client Self-Selection 
}
\author{

\IEEEauthorblockN{Kaan Okumus$^{*}$, Khac-Hoang Ngo$^{\dagger}$, Unnikrishnan Kunnath Ganesan$^{*}$, \\Giuseppe Durisi$^{*}$, 
Erik G. Str\"om$^{*}$, and Shashi Raj Pandey$^{\ddagger}$
}
\IEEEauthorblockA{
$^{*}$Department of Electrical Engineering, Chalmers University of Technology, 41296 Gothenburg, Sweden\\ 
$^{\dagger}$Department of Electrical Engineering, Link\"oping University, 58183 Link\"oping, Sweden\\ 
$^{\ddagger}$Department of Electronic Systems, Aalborg University, 9220 Aalborg, Denmark
\vspace{-1em}
\thanks{This work was supported in part by the Swedish Research Council under grants 2021-04970 and 2022-04471, and by the Swedish Foundation for Strategic Research. The work of K.-H. Ngo was supported in part by the Excellence Center at Linköping--Lund in Information Technology (ELLIIT). The work of S.R. Pandey was supported in part by DFF-Forskningsprojekt1 “NETML” with grant no. 4286-00278B.}
}
}

\date{\today}

\begin{document}

\maketitle

\begin{abstract}
	We address the client-selection problem in federated learning  over wireless networks under data heterogeneity. 
	Existing client-selection methods often rely on server-side knowledge of client-specific information, thus compromising privacy. To overcome this issue, we propose a client self-selection strategy  
    based solely on the comparison between locally computed training losses and a centrally updated selection threshold. Furthermore, to support robust aggregation of clients' updates over wireless channels, we integrate this client self-selection strategy into the recently proposed type-based unsourced multiple-access framework over distributed multiple-input multiple-output (D\nobreakdash-MIMO) networks.  
	The resulting scheme is completely unsourced: the server does not need to know the identity of the clients. Moreover, no channel state information 
	is required, neither at the clients nor at the server side.
	Simulation results conducted over a D\nobreakdash-MIMO wireless network 
    show that the proposed self-selection strategy matches the performance of a comparable state-of-the-art server-side selection method and consistently outperforms random client selection. 
\end{abstract}

\section{Introduction}

Future wireless networks are envisioned to support intelligent services through a massive number of connected devices. One key application is over-the-air federated learning, where multiple devices, also called clients, collaboratively train a model under the coordination of a central server, while preserving privacy. 
In each training round, clients receive the global model, perform local training, and transmit their updates to the server over a shared wireless medium. These transmissions are  sporadic due to limited communication resources and intermittent connectivity. Federated averaging (FedAvg)~\cite{mcmahan17} is a foundational method that performs these steps iteratively over multiple rounds.

When the number of clients is large, selecting a subset of them in each round becomes essential to reduce communication and computation overhead. Random client selection is simple and widely used. However, it becomes inefficient under data heterogeneity, because it results in  
``client drift'': as the training rounds progress, each local update is pulled toward a local optimum, leading to slow and unstable convergence~\cite{scaffold_20}. 
It also entails a participation gap~\cite{yuan_2022_what}, which 
captures the distribution shift between the data observed during training and the global data distribution.

The SCAFFOLD algorithm~\cite{scaffold_20} mitigates client drift using control variates (e.g., local and aggregated gradients) exchanged between clients and the server.
However, this method doubles the communication cost per round, and its performance degrades with sporadic participation, since local control variates often become stale~\cite{reddi_21}.
The participation gap can be estimated if all clients share held-out data with the server.
This enables client-selection algorithms that improve convergence~\cite{singhal_24}.
However, sharing held-out data violates client privacy.
The power-of-choice (PoC) strategy~\cite{cho_22} partially alleviates this issue by letting the server select clients based on reported training losses.
However, these losses still expose client-specific information. The design of a truly privacy-preserving client-selection mechanism remains an open challenge.

In all works reviewed so far, the communication channel between the clients and the server is assumed error-free. Existing works on over-the-air federated learning mainly rely on analog over-the-air computation (AirComp) techniques, e.g.,~\cite{amiri_20, zhu_20, yang_20, jeon_21}, which exploit waveform superposition for model aggregation but require strict synchronization, channel pre-equalization, and specialized analog hardware.  
Digital AirComp methods, e.g.,~\cite{zhu_21_one_bit_AirComp, sahin_23, sahin_24_balanced_num}, retain digital modulation but remain limited to low-resolution scalar quantization, 
and thus scale poorly with the model dimension and the number of clients. 
To overcome these limitations, Qiao et \textit{al.}~\cite{qiao_gunduz_fl} introduced the massive digital (MD)-AirComp framework, where local update transmission and aggregation are formulated as an unsourced multiple access (UMA)~\cite{polyanskiy2017} problem. In this framework, all clients employ the same encoder, and the server needs to recover the set of (quantized) local updates. 
Specifically, it needs to estimate the number of distinct local updates and their
multiplicities, i.e., their \emph{type}.
This task falls under the framework of type-based
UMA (TUMA)~\cite{ngo2024_tuma}. In MD-AirComp~\cite{qiao_gunduz_fl}, type
estimation is performed by an approximate message passing (AMP) decoder that
requires channel state information (CSI) at the clients for channel pre-equalization.
A CSI-free AMP-based type estimator was developed in~\cite{tuma_fading_2025}
for communication over a distributed multiple-input multiple-output (D-MIMO) network.

\subsubsection*{Contributions}
In this paper, we address the joint challenge of privacy-preserving client
selection and reliable local-update transmission in federated learning over
a D-MIMO network, under heterogeneous data and sporadic client activity.
We first propose a client self-selection strategy in which each client
autonomously decides its participation by comparing its local training loss with
a selection threshold, broadcast by the server.
The server updates this threshold according to the
estimated number of participants, rather than any client-specific information.
We then integrate this client-selection strategy into the CSI-free TUMA
framework~\cite{tuma_fading_2025} for model aggregation over a D-MIMO network.
The resulting scheme is completely unsourced: no information about the clients' identity is available at the server.
Numerical results show that our CSI-free TUMA scheme achieves a test accuracy
comparable to that attainable under ideal communication and superior to that of
MD-AirComp.
Furthermore, our proposed client self-selection
method achieves convergence performance close to that of PoC~\cite{cho_22} and
consistently outperforms random client selection.

\subsubsection*{Notation}
Sets are denoted by calligraphic letters (e.g., $\setS$),
vectors by bold lowercase letters (e.g., $\vecx$), and matrices by bold uppercase letters (e.g., $\matX$). We denote the Kronecker product by $\otimes$, let $[n]=\set{1,\dots,n}$, and use $\diag(x_1, \dots, x_n)$ to indicate a diagonal matrix with diagonal entries $x_1, \dots, x_n$. 

\section{System Model} \label{sec:unsourced_FL_sys_model}
We consider a federated learning setup consisting of $K$ clients and a single server. Each client $k\!\in\![K]$ holds a local dataset~$\mathcal{B}_k$ of size $N_k\!=\!|\mathcal{B}_k|$. The objective is to jointly train a global model~$\vecw\!\in\!\mathbb{R}^{W}$ by minimizing the global empirical~loss
\begin{equation}
	\vecw^\star =
	\argmin_{\vecw \in \mathbb{R}^{W}} \sum_{k=1}^{K}\frac{N_k}{N_{\text{tot}}}
	f_k(\vecw;\mathcal{B}_k), \label{main_min_prob}
\end{equation}
where $N_{\text{tot}}=\sum_{k=1}^{K} N_k$, and $f_k(\vecw; \setB_k)$ denotes the local loss of client~$k$. As a baseline, we follow the FedAvg algorithm~\cite{mcmahan17}, while extending it to an unsourced communication setting as described next. 

Training proceeds over rounds $t \in [T]$. We assume that, at the beginning of each round~$t$, each client is \emph{activated} independently with probability~$\lambda$. The server broadcasts the current global model~$\vecw^{(t-1)}$ to all active clients through an ideal downlink channel. Each active client then decides whether to participate in the round (and become \textit{selected}) based on a local selection strategy. The resulting set of participating clients is denoted by~$\mathcal{S}^{(t)}\!\subseteq\![K]$ and its cardinality by $L^{(t)}$. 

Each selected client~$k\!\in\!\mathcal{S}^{(t)}$ performs local training for~$E$ epochs with mini-batch size~$V$ and learning rate~$\eta$, using mini-batch stochastic gradient descent:
\begin{equation}
	\vecw_{k,\ell}^{(t)}
	= \vecw_{k,\ell-1}^{(t)}
	- \frac{\eta}{V}
	\sum_{n\in\setE}
	\nabla f_k(\vecw_{k,\ell-1}^{(t)};n),
	\quad \ell\!\in\![E],
\end{equation}
where $\setE\!\subset\!\mathcal{B}_k$ is a randomly sampled mini-batch and
$\vecw_{k,0}^{(t)}=\vecw^{(t-1)}$.
After $E$ epochs, client~$k$ obtains the local model $\vecw_k^{(t)}=\vecw_{k,E}^{(t)}$, and computes the local model update
\begin{equation}
	\vecDelta_k^{(t)} = \vecw_k^{(t)} - \vecw^{(t-1)}.
\end{equation}
Client~$k$ then encodes $\vecDelta_k^{(t)}$ into a codeword and transmits this codeword over the wireless channel.

The server receives a signal $\matY^{(t)}$ and uses it to obtain an estimate $[\,\widehat{\vecDelta}_1^{(t)}, \ldots,
	\widehat{\vecDelta}_{\widehat{L}^{(t)}}^{(t)}\,]$ of the transmitted local updates.
Here, $\widehat{L}^{(t)}$ is the server-side estimate of $L^{(t)}$. The server then

\noindent
updates the global model as follows:\footnote{Note that, since
	our transmission strategy (described in Section~\ref{sec:quant-tuma-cf}) ensures that clients' identities remain unknown to the server, the dataset-size-weighted averaging used in~\eqref{main_min_prob} cannot be applied to~\eqref{eqn:ufl-global-agg}.}
\begin{equation}
	\vecw^{(t)} =
	\vecw^{(t-1)} + \eta\sub{g}\, \frac{1}{\widehat{L}^{(t)}}
	\sum_{i=1}^{\widehat{L}^{(t)}}
	\widehat{\vecDelta}_i^{(t)}, \label{eqn:ufl-global-agg}
\end{equation}
where $\eta\sub{g}$ is the global learning rate.

Our broad goal is to design a local client-selection strategy as well as a communication scheme for which $\vecw^{(t)}$ converges rapidly to $\vecw^\star$.

\section{The Proposed Client Self-Selection Strategy} \label{sec-self-select}
In this section, we propose a client-selection algorithm and evaluate its performance under error-free communication.
We will then investigate the performance attainable under a realistic D-MIMO deployment in Section~\ref{sec:quant-tuma-cf}.

\subsection{Self-Selection Algorithm}
We propose a client self-selection mechanism that mimics the behavior of the PoC selection rule~\cite{cho_22}.  
Unlike PoC though, where the server selects clients based on reported utilities, here each active client decides autonomously whether to participate, without revealing any client-specific information to the server.
Specifically, each client determines its participation by comparing its local training loss with a threshold broadcast by the server. The server updates the threshold to adjust the expected number of participants. 
A detailed description of the proposed strategy is provided next.

At the beginning of round~$t$, each active client receives the current selection threshold $\theta^{(t)}$ and the global model $\vecw^{(t-1)}$ broadcast by the server.

\subsubsection{Candidate set formation}
Each active client enters the candidate set with probability $p_{\text{cand}}=d/(\lambda K)$, so that, on average, the candidate set contains $d$ clients. This ensures that decisions are based on a diverse subset of clients across each round, 
which prevents overfitting, as argued in~\cite{cho_22}.

\subsubsection{Local evaluation and self-selection}
Each candidate computes its local loss on the received global model,
\begin{equation}
	\ell_k^{(t)} = f_k(\vecw^{(t-1)};\mathcal{B}_k),
\end{equation}
and decides to participate with probability
\begin{equation}
	\Pr\{k \in \setS^{(t)} \mid \ell_k^{(t)}\} = \sigma\!\left(a \big(\ell_k^{(t)} - \theta^{(t)}\big)\right), \label{eqn-prob_sel_sigmoid}
\end{equation}
where the sigmoid function $\sigma(x) = 1/(1+e^{-x})$ is used to convert the local-loss  improvement $\ell_k^{(t)} - \theta^{(t)}$ into a participation probability, and $a>0$ controls the steepness of the transition between $0$ and $1$ probability values. Specifically, 
a smaller $a$ introduces randomness that helps prevent overfitting by avoiding the repeated selection of the same high-loss clients.

\subsubsection{Threshold update} After aggregation, the server updates the threshold to maintain the average number of participating clients per round close to a target value $K_{\text{tar}}$: 
\begin{equation}
	\theta^{(t+1)} = \theta^{(t)} + \xi(\widehat{L}^{\,(t)}-K_{\text{tar}}). \label{threshold_update}
\end{equation}
Here, $\xi>0$ is a step-size parameter. An estimated participation $\widehat{L}^{\,(t)}$ higher than $K_{\text{tar}}$ increases the threshold, thereby reducing future participation probabilities. 

The overall strategy is summarized in Algorithm~\ref{algo-self_select}. The client-side computational overhead of this strategy is negligible relative to local training, as it requires only a single loss evaluation and a scalar threshold comparison per round.

\begin{algorithm}[!t]
    \caption{Decentralized Client Self-Selection}
    \label{algo-self_select}
    \begin{algorithmic}[1]
        \State Server: broadcast $(\vecw^{(t-1)},\theta^{(t)})$
        \For{each active client $k$}
        \State join candidate set with probability\ $p_{\text{cand}}=d/(\lambda K)$
        \If{candidate}
        \State compute local loss $\ell_k^{(t)} = f_k(\vecw^{(t-1)};\mathcal{B}_k)$
        \State self-select with probability \ $\sigma(a(\ell_k^{(t)}-\theta^{(t)}))$
        \If{selected} do local training and transmit $\vecDelta_k^{(t)}$ \EndIf
        \EndIf
        \EndFor
        \State \textbf{Server:} estimate $\widehat{L}^{(t)}$ and update $\theta^{(t+1)}$ via~\eqref{threshold_update} 
    \end{algorithmic}
\end{algorithm}

\subsection{Performance Evaluation Under Perfect Communication} \label{sec:perfect_sim}
We evaluate the proposed self-selection strategy under an error-free
communication scenario, in which the local updates are delivered to the
server without errors. 
We consider the FMNIST dataset, which consists of $\num{60000}$ samples, divided into
training, validation, and test sets with ratios $0.8$, $0.1$, and $0.1$,
respectively. The model is a multilayer perceptron  
with an input dimension of $784$, corresponding to a flattened $28\times28$ FMNIST image. It includes two fully connected hidden layers of sizes $64$ and $30$, respectively, each followed by a ReLU
activation. This results in a model dimension $W=\num{52500}$. A dropout layer with rate~$0.5$ is inserted between the second
hidden layer and the output layer to improve generalization. 
A $\text{LogSoftmax}$ activation is applied to the output of the last layer. The loss function is a negative log-likelihood. Each selected client performs local training for $E=30$ epochs per round with a mini-batch size $V=64$, local learning rate $\eta=0.001$, and global learning rate $\eta\sub{g}=1$. Data are randomly distributed among clients using a Dirichlet distribution with parameter $\alpha=2$, resulting in moderately heterogeneous local datasets. 
We set the total number of clients to $K=1000$ and the activation probability to
$\lambda=0.8$. Following~\cite{cho_22}, we choose a participation ratio $K_\text{tar}/K=0.1$, which results in $K_{\text{tar}}=100$. Training is conducted for $T=500$ rounds.

As a benchmark, we consider two alternative strategies:
\begin{itemize}
	\item \textbf{Random:} each active client independently selects itself with probability $K_{\text{tar}}/(\lambda K)$.
	\item \textbf{PoC:} selection is performed at the server side, based on the highest reported local losses~\cite{cho_22}. We set the size of the candidate set to $d=2\times K_\text{tar}=200$.
\end{itemize}
We compare these two strategies with the proposed self-selection strategy, for the case where we set $a=50$   in~\eqref{eqn-prob_sel_sigmoid} and  $\theta^{(1)}=2.32$, $\xi=0.004$ in~\eqref{threshold_update}.\footnote{The parameters $a,\theta^{(1)},$ and $\xi$ are selected through empirical hyperparameter tuning.} 
Similar to PoC, we set $d = 200$, and thus $p_\text{cand}=d/(\lambda K)=0.25$.

\begin{figure}[t!]
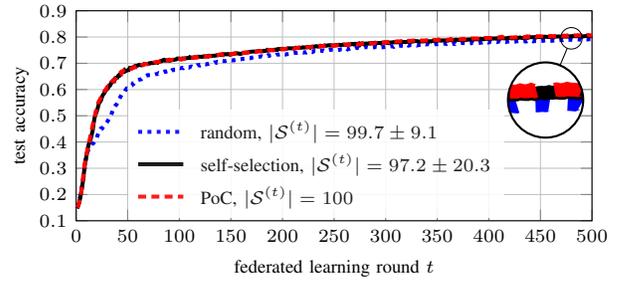

	\centering
	\include{figs/test_acc_vs_round_perfect_novq_Ktar100_K1000.tex}
	\vspace{-1cm}
	\caption{Test accuracy as a function of the number $t$ of federated learning rounds for different selection strategies resulting in approximately the same average number of participating clients $K_\text{tar}=100$, under error-free communication. The actual average number of participating clients $|\setS^{(t)}|$ for each selection strategy, as well as its standard deviation are reported in the legend.}
	\label{fig:conv_perfect_comm}
	\vspace{-.3cm}
\end{figure}

Fig.~\ref{fig:conv_perfect_comm} shows the test accuracy as a function of the federated learning rounds for the three selection strategies under error-free communication. The proposed self-selection strategy exhibits convergence behavior nearly identical to that of PoC, while outperforming random selection. This demonstrates that client self-selection can match
the performance of server-side selection without compromising privacy.

\section{Type-Based Unsourced Federated Learning} \label{sec:quant-tuma-cf}
In this section, we address the transmission of the local model updates over a wireless channel.
Specifically, we consider a D-MIMO scenario, in which all participating users transmit their updates simultaneously, and a TUMA decoder~\cite{tuma_fading_2025} is used to estimate the quantized local updates.

\subsection{Quantization of Local Updates} \label{subsec-quant}
Following~\cite{qiao_gunduz_fl, amiri_error_acc_w_quant}, we assume that each client transmits a quantized version of its local model update $\vecDelta_k^{(t)}$, obtained through vector quantization with error accumulation. Let $\setQ = \{\vecq_1,\ldots,\vecq_M\}$ denote the quantization codebook, where each $\vecq_m \in \mathbb{R}^Q$ represents a quantization codeword of dimension~$Q$. Note that $Q=1$ corresponds to scalar quantization. The number of quantization bits is $J = \log_2M$.
At round~$t$, quantization proceeds as follows.
Let $\vece_k^{(t)}$ denote the accumulated quantization error for client~$k$ in round $t$, initialized as
$\vece_k^{(0)}=\veczero$.
The client first incorporates the accumulated quantization error into the model update by computing
\begin{equation}
	\Bar{\vecs}_k^{(t)} = \vecDelta_k^{(t)} + \vece_k^{(t-1)}.
\end{equation}
Then, $\Bar{\vecs}_{k}^{(t)}$ is divided into
$D=\lceil W/Q\rceil$ contiguous subvectors\footnote{Zero-padding is applied to $\Bar{\vecs}_{k}^{(t)}$ before this division when
	$Q$ does not exactly divide $W$, ensuring that all $D$ subvectors have
	equal length~$Q$.}
$\{\Bar{\vecs}_{k}^{(d,t)}\}_{
d=1}^{D}$, and each one is quantized  to its nearest codeword according to
\begin{equation}
	M_{k}^{(d,t)} = \argmin_{m\in[M]}\!
	\left\lVert \vecq_m - \Bar{\vecs}_{k}^{(d,t)} \right\rVert_2,
	\quad d\in[D]. \label{eq-quantfl-3}
\end{equation}
The quantized subvectors are concatenated back into a vector\footnote{Truncation is performed if zero padding was necessary.} of length $W$ as follows:
\begin{equation}
	\vecs_k^{(t)} =
	\mathrm{concat}\!\left(\vecq_{M_{k}^{(1,t)}}, \ldots,
	\vecq_{M_{k}^{(D,t)}}\right). \label{eq-quantfl-4}
\end{equation}
Finally, the quantization error to be carried over to the next round is updated as
\begin{equation}
	\vece_k^{(t)} = \Bar{\vecs}_k^{(t)} - \vecs_k^{(t)}. \label{eq-quantfl-5}
\end{equation}

Each client thus produces a sequence of quantization indices~$\{M_{k}^{(d,t)}\}_{d=1}^{D}$, which are transmitted over $D$ subrounds.
Specifically, these indices are mapped to communication codewords according to the TUMA coding scheme described next.

\subsection{TUMA over D-MIMO}

We consider a D-MIMO network where $B$ access points (APs), each equipped with $A$ antennas (yielding $F= A\times B$ antennas in total) are connected to the server via fronthaul links. The APs jointly serve single-antenna clients, randomly distributed within a coverage area $\setD$.
This area is
partitioned into $U$ non-overlapping zones
$\{\mathcal{D}_u\}_{u=1}^{U}$. Each zone consists of locations with similar large-scale fading coefficients (LSFCs), so that users within the same zone experience approximately the same average path loss towards the APs. This partitioning mitigates the effect of path loss variability~\cite{cakmak_2025_journal}.
We denote the position of each AP~$b$ by $\nu_b \in \setD$.

All clients use the same codebook $\matC\in \mathbb{C}^{N \times \adjustedbar{M}}$, where $N$ is the blocklength and $\adjustedbar{M}=U \times M$ is the total number of codewords. The codebook is evenly partitioned into zone-specific subcodebooks: $\matC=[\matC_1, \ldots, \matC_U]$, where $\matC_u=[\vecc_{u,1},\ldots,\vecc_{u,M}]$, and each codeword satisfies the average-power constraint $\lVert \vecc_{u,m}\rVert_2^2 \leq 1$.
The mapping from quantization index to communication codeword is one-to-one and zone dependent:  client $k$ in zone $u$ maps the quantization index $M_{k}^{(d,t)}\!\in\![M]$ 
to codeword $\vecc_{u,M_{k}^{(d,t)}}$, to be transmitted in subround $d$ of round~$t$.

We denote the number of clients transmitting codeword~$m$ in zone~$u$ during subround~$d$ of round~$t$ by $k_{u,m}^{(d,t)}$.
Furthermore,
we let $\veck_u^{(d,t)}=[k_{u,1}^{(d,t)},\ldots,k_{u,M}^{(d,t)}]^{\top}$ denote the corresponding zone-based multiplicity vector, and
$\veck^{(d,t)}=[k_{1}^{(d,t)},\ldots,k_{M}^{(d,t)}]^{\top}$ with $k_m^{(d,t)} = \sum_{u=1}^U k_{u,m}^{(d,t)}$, denote the global multiplicity vector.
Note that $\lVert \veck^{(d,t)} \rVert_1$ is the number of participating clients in round~$t$.
The type $\vect^{(d,t)}$ is then obtained as the vector of normalized multiplicities, i.e., $\vect^{(d,t)}=\veck^{(d,t)}/\lVert \veck^{(d,t)} \rVert_1$.
This vector represents the empirical distribution of transmitted quantized updates across all participating clients. 

We index by $(u,k)$ the $k$th participating client in zone~$u$. The channel between the client~$(u,k)$ and the AP~$b$ is modeled as a spatially-white quasi-static Rayleigh fading channel. Specifically, the channel coefficients are assumed to be independent across antennas, APs, and clients and constant over each codeword. The channel vector $\vech_{u,k}^{(d,t)} \in \mathbb{C}^F$ between the client~$(u,k)$ and all receive antennas is distributed as $\mathcal{C}\mathcal{N}(\veczero, \matSigma(\rho_{u,k}))$, where $\rho_{u,k}$ denotes the position of user~$(u,k)$ and $\matSigma(\rho_{u,k}) = \diag(\gamma_1(\rho_{u,k}), \ldots, \gamma_{B}(\rho_{u,k})) \otimes \matidentity_{A}$ with $\gamma_b(\rho_{u,k})$ being the position-specific LSFC. When multiple users transmit the same codeword, their contributions are superimposed at the receiver. For a codeword~$\vecc_{u,m}$, the effective channel vector is given by
\begin{equation}
	\vecx_{u,m}^{(d,t)} =
	\begin{cases}
		\sum_{(u,k) \sothat M_{k}^{(d,t)}=m} \vech_{u,k}^{(d,t)} & \text{if } k_{u,m}^{(d,t)}>0, \\
		\veczero                                                 & \text{if } k_{u,m}^{(d,t)}=0.
	\end{cases} \label{expression_x}
\end{equation}
We define the effective channel matrix $\matX^{(d,t)}_u$, which collects the effective channel vectors for all codewords in zone $u$, as $\matX^{(d,t)}_u = [\vecx_{u,1}^{(d,t)}, \ldots, \vecx_{u,{M}}^{(d,t)}]^\top \in \mathbb{C}^{{M} \times F}$. The aggregated received signal across all APs is then given by
\begin{align} \label{eqn-mainreceivedsignal}
	\matY^{(d,t)}
	= \sqrt{NP}  \sum_{u=1}^{U} \matC_u \matX_u^{(d,t)}  + \matW^{(d,t)},
\end{align}
where $\matW^{(d,t)}$ is the additive white Gaussian noise whose elements are drawn independently from a $\mathcal{C}\mathcal{N}(0,\sigma_w^2)$ distribution, and $P$ denotes the average transmit power.

The received matrix $\matY^{(d,t)}$ serves as the input to
the TUMA decoder. 
The decoder operates over multiple subrounds $d\in [D]$, and provides in each subround an estimate  $\widehat{\vect}^{(d,t)}$ of the type of the quantized local updates
transmitted in that subround. 
Using the estimated types, the server performs a global model update for subblock $d$ as (see~\eqref{eqn:ufl-global-agg})
\begin{equation}
	\vecw_d^{(t)}
	=\vecw_d^{(t-1)}+\eta\sub{g} \sum_{m=1}^{M} \widehat{t}^{(d,t)}_{m}\, \vecq_m.\label{eq:block_update_from_type}
\end{equation}
After processing all subrounds, the subblock-based updated global models $\{\vecw_d^{(t)}\}_{d=1}^{D}$ are concatenated (and, possibly, truncated to length $W$) to obtain  the updated global model~$\vecw^{(t)}$.

As a TUMA decoder, we adopt the centralized multisource AMP algorithm described in~\cite[Algo.~1]{tuma_fading_2025}, which iteratively estimates the multiplicity vector
$\widehat{\veck}_u^{(d,t)}$ using a Bayesian denoiser. In the absence of prior
knowledge on client data distributions or participation statistics, we employ a truncated Poisson prior
\begin{equation}
	k_{u,m} \sim \mathrm{Pois}_{[0:K_{\text{max}}]}\!\left(
	\frac{K_{\text{tar}}}{U \times M}\right),
	\quad K_{\text{max}} = \left\lceil \frac{K_{\text{tar}}}{U}\right\rceil, \label{eq:prior}
\end{equation}
where $\mathrm{Pois}_{[0,a]}(\alpha)$ denotes the Poisson distribution with mean~$\alpha$, truncated to the interval $[0,a]$.
This prior promotes sparsity by reflecting that only a limited number of codewords are active in each zone during a given round. 
We refer the reader to~\cite[Sec.~3]{tuma_fading_2025} for more details on the 
TUMA decoder.

\section{Performance over D-MIMO Networks}

We now integrate the client-selection strategy proposed in Section~\ref{sec-self-select} with the TUMA-based model aggregation over a D-MIMO network described in Section~\ref{sec:quant-tuma-cf}, and evaluate the resulting test-accuracy for the same federated learning problem considered in Section~\ref{sec:perfect_sim}. 

To construct the vector quantization codebook, we follow the procedure described in~\cite{qiao_gunduz_fl}. Specifically, we assume that the server has a separate dataset, used solely for
updating the quantization codebook $\setQ$.\footnote{The presence of this dataset does not violate privacy, as it is different from the clients' local datasets.} At each round, the server performs one local training step and obtains a local update with error accumulation, which is then divided into $D$ subvectors of dimension~$Q$. Using these subvectors as samples, the server updates the quantization codebook using the K\nobreakdash-means++ algorithm~\cite{arthur_kmeans}. Specifically,  the server finds $M=2^J$ cluster centroids and uses them as quantization codewords. The updated codebook is broadcast to the clients together with the global model.

\begin{figure}[t!]
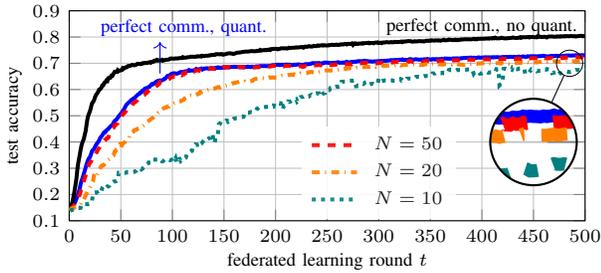

	\centering
	\include{figs/test_acc_vs_round_comm_different_N_self}
	\vspace{-1cm}
	\caption{Test accuracy as a function of the number $t$ of federated learning rounds for the TUMA decoder with client self-selection; we consider vector quantization with parameters $J=7$, $Q=30$ and for different transmission blocklength values $N \in \{10, 20, 50\}$.}
	\label{fig:test_acc_N}
	\vspace{-.3cm}
\end{figure}

We consider a simple D-MIMO deployment resulting in a $3 \times 3$ square grid layout, with each square having a side of length $\qty{100}{m}$.
Specifically, we assume that $B=40$ APs, each equipped with $A=4$ antennas, are evenly placed along the square boundaries.
This leads to $F=160$ receive antennas in total.
Similar to~\cite{cakmak_2025_journal}, the LSFC is modeled as $\gamma_b(\rho) = 1/(1 + \left( |\rho - \nu_b|/d_0\right)^\alpha)$, where $\alpha=3.67$ and $d_0=\qty{13.57}{m}$.
The coverage area is divided into $U=9$ zones, each one corresponding to one of the squares.   
We define the received SNR as $\text{SNR}_{\text{rx}} = \text{SNR}_{\text{tx}} / (1 + (\varsigma/d_0)^\alpha)$, where $\varsigma=\qty{50}{m}$ denotes the distance between a zone centroid and its closest AP. In all simulations, we fix $\text{SNR}_{\text{rx}} = \qty{10}{dB}$ and the average per-symbol transmit power for communication at $P = \qty{1}{mW}$. The codewords~$\{\vecc_{u,m}\}$ are independently generated by first sampling each entry independently from a Gaussian distribution~$\mathcal{C}\mathcal{N}(0, 1/N)$, and then normalizing each codeword such that $\lVert \vecc_{u,m}\rVert_2^2=1$. 
We use the truncated Poisson prior in~\eqref{eq:prior} with $K_\text{max}=8$, and adopt the learning hyperparameters specified in Section~\ref{sec:perfect_sim}.\footnote{The code used to obtain all results is available at \url{https://github.com/okumuskaan/ufl_tuma} to support reproducible research.}

In Fig.~\ref{fig:test_acc_N}, we show the test accuracy across federated learning rounds, obtained with the TUMA decoder and our proposed self-selection client strategy, for three different communication blocklengths, i.e., $N\!\in\!\{10,20,50\}$.
We set $J\!=\!7$ and $Q\!=\!30$.\footnote{Lower values of $Q$ and higher values of $J$ yield finer quantization but increase communication overhead and decrease communication performance, since a smaller $Q$ leads to more communication subrounds and a larger $J$ increases the number of transmitted bits and, hence, the transmission rate.
}
We also include for reference the curves for the case of perfect communication, both with and without vector quantization, to isolate the performance loss due to vector quantization with the chosen parameters.
We see from the figure that, as the blocklength $N$ increases, the test accuracy with TUMA improves and approaches that of perfect communication.
Indeed, longer blocklengths result in lower rates, and, hence, fewer type-estimation errors.
Notably, the system operates in the regime $N \ll 2^{J}=128$, making the use of an orthogonal codebook infeasible. 

\begin{figure}[t!]
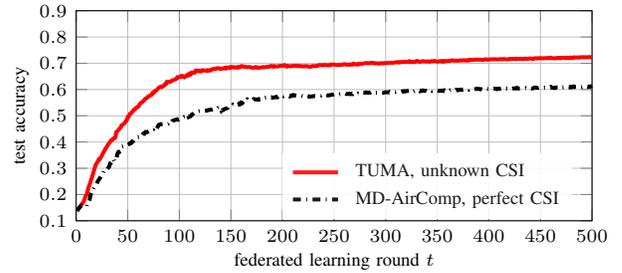

	\centering
	\include{figs/test_acc_vs_round_different_decoder_self_v2}
	\vspace{-1cm}
	\caption{Test accuracy as a function of the number $t$ of federated learning rounds; we compare TUMA and MD-AirComp with perfect CSI for $N=50$, $J=7$, $Q=30$, and client self-selection.}
	\label{fig:test_acc_dec}
	\vspace{-.4cm}
\end{figure}

In Fig.~\ref{fig:test_acc_dec}, we compare the test accuracy achieved with the TUMA decoder against the one attained by an extension of the MD-AirComp scheme proposed in~\cite{qiao_gunduz_fl} to the D-MIMO setup.
We set $N=50$. For MD-AirComp, 
we assume perfect CSI\footnote{Recall that the TUMA coding scheme, instead, does not require any instantaneous CSI.} and perform transmitter-side channel pre-equalization following~\cite{qiao_gunduz_fl}. Furthermore, we normalize the transmitted signal so that the received SNR remains fixed at $\text{SNR}_{\text{rx}}=\qty{10}{dB}$, ensuring a fair comparison with TUMA.
As shown in Fig.~\ref{fig:test_acc_dec}, MD-AirComp suffers a significant performance degradation. 

\begin{table*}[t!]
	\centering
	\caption{End-to-End Performance for Different Scenarios}
	\label{tab:fl_summary_results}
	\setlength{\tabcolsep}{6pt}
	\renewcommand{\arraystretch}{1.4}
	\scriptsize
	\begin{tabular}{lccccccc}
		\hline
		\textbf{Scenario / Coding Scheme}      & \textbf{$(J,Q)$}               & \textbf{$N$} & \textbf{Selection} &
		\textbf{Final Test Accuracy (\%)}          & \textbf{Number of Rounds to $70\%$} &
		$|\setS^{(t)}|$                                                                                                                                \\
		\hline
		Perfect communication, no quantization & --                             & $\infty$     & Self               & $80.5$ & $73$  & $97.2\pm 20.3$  \\
		Perfect communication, quantization    & $(7, 30)$                      & $\infty$     & Self               & $73.1$ & $236$ & $97.2 \pm 21.6$ \\
		\hline
		\textbf{TUMA}                          & $(7, 30)$                      & $50$         & \textbf{Self}      & $72.3$ & $295$ & $94.4\pm34.1$   \\
		TUMA                                   & $(7,30)$                       & $50$         & PoC                & $72.8$ & $284$ & 100             \\
		TUMA                                   & $(7,30)$                       & $50$         & Random             & $70.0$ & $478$ & $99.7\pm9.1$    \\
		MD-AirComp & $(7, 30)$                      & $50$         & Self               & $61.4$     & --    & $89.6\pm 43.4$              \\
        TUMA                                   & $(5, 30)$                      & $50$         & Self               & $70.1$ & $466$ & $94.3\pm 35.3$  \\
		TUMA                                   & $(7, 30)$                      & $20$         & Self               & $70.9$ & $430$    & $90.1\pm 40.3$ \\
		\hline
	\end{tabular}
	\vspace{-5mm}
\end{table*}

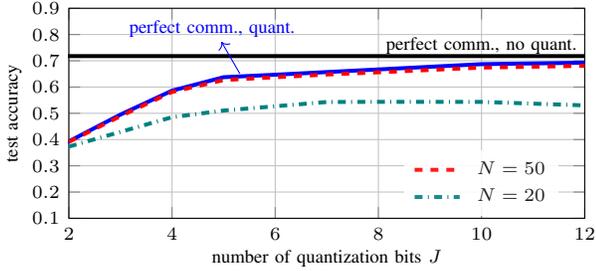
\begin{figure}[t!]
	\centering
	\begin{tikzpicture}[
]
        \tikzstyle{every node}=[font=\scriptsize] 
        \begin{axis}[
            scale only axis,
            width=2.7in,
            height=1.1in,
            grid=both,
            grid style={line width=.1pt, draw=gray!10},
            major grid style={line width=.2pt,draw=gray!50},
            xlabel={\scriptsize number of quantization bits $J$},
            ymin=0.1,
            ymax=0.9,
            xmin=2,
            xmax=12,
            ylabel={\scriptsize test accuracy},
            ylabel style = {yshift=-1mm},
            xlabel style = {yshift=1mm},
            ytick = {0.1,0.2,0.3,0.4,0.5,0.6,0.7,0.8, 0.9},
            xtick = {0,2,4,6,8,10,12},
            legend cell align={left},
            legend style={draw=none,opacity=.9,at=
            {(0.65,0.009)},anchor=south west, nodes={scale=1, transform shape}, 
            },
        ]

            \addplot[ color=black, line width = 1.5pt, forget plot]
            table[row sep=crcr]{
	-1	0.7179	\\
    13  0.7179  \\
            };

            \addplot[color=blue, line width = 1.5pt, forget plot]
            table[row sep=crcr]{
    1   0.328    \\
    2   0.393   \\
    3   0.4950  \\
    4   0.5862  \\
    5   0.6375  \\
    6   0.6470  \\
	7	0.6571	\\
    10  0.6872  \\
    12  0.6928  \\
            }
                node [blue, pos=.4, inner sep=0, yshift=18pt, xshift=-12pt, align=center, scale=1] {\scriptsize perfect comm., quant.}
                node [blue, pos=.4, inner sep=0, yshift=7pt, xshift=-5pt, rotate=30] {\tikz{\draw[->, blue, line width=0.4pt] (0,0) -- (0,0.14);}}
            ;

            \addplot[color=red, dashed, line width = 1.5pt]
            table[row sep=crcr]{
    2   0.3912  \\
    3   0.4890  \\
    4   0.5812  \\
    5   0.6267  \\
	7	0.6478	\\
    8   0.6561  \\
    10  0.6742  \\
    12  0.6808  \\
            };
            \addlegendentry{\scriptsize $\,$ $N=50$}

            \addplot[color=teal, dashdotted,line width = 1.5pt]
            table[row sep=crcr]{
    2   0.3731  \\
    4   0.4851  \\
    5   0.5105  \\
	7	0.543 \\
    8   0.5436  \\
    10  0.5434  \\
    12  0.5301  \\
            };
            \addlegendentry{\scriptsize $\,$ $N=20$}

        \node[align = center,color=black, scale=1] at (axis cs:10.0,0.76) () {\scriptsize perfect comm., no quant.};

        \end{axis}
\end{tikzpicture}%
	\vspace{-1cm}
	\caption{Test accuracy achieved at round $t=100$ versus the number of quantization bits $J$ for TUMA decoder and client self-selection; $Q=30$.}
	\label{fig:tradeoff_quant_comm}
	\vspace{-.3cm}
\end{figure}

In Fig.~\ref{fig:tradeoff_quant_comm}, we show the test accuracy after~$100$ rounds as a function of the number of quantization bits~$J$ for
the proposed client self-selection method, when $N\in\{20,50\}$ and $Q=30$. 
For $N=50$, the test accuracy increases monotonically with~$J$, for the values of $J$
considered in the figure.
Instead, the behavior is not monotonic for $N=20$, and the test accuracy
peaks at $J=8$.
This behavior can be explained by the tradeoff between quantization distortion
(decreasing when $J$ increases) and
communication errors (increasing when $J$ increases).
Note that this optimum operating point occurs at a value of $J$ that is much larger than
the one compatible with the use of an orthogonal codebook.

In Table~\ref{tab:fl_summary_results}, we summarize the end-to-end performance of the proposed framework under different communication conditions and client-selection strategies. For $N=50$, the proposed self-selection strategy outperforms random selection, especially in terms of convergence speed, and achieves performance close to PoC, confirming its effectiveness in preserving privacy without sacrificing accuracy.

\section{Conclusion}
We proposed a client self-selection strategy for over-the-air federated learning, where clients decide their participation based on locally computed training loss and a centrally updated threshold. We integrated this strategy with TUMA-based digital over-the-air aggregation over a D\nobreakdash-MIMO wireless network.  The resulting scheme does not require CSI,
and operates according to an unsourced communication paradigm, in which
the server does not need to know the identity of the clients. Our simulation results show that the proposed method achieves a test accuracy that matches that of PoC~\cite{cho_22}---a state-of-the-art server-side selection method based on reported training losses. 
Optimizing the system parameters to maximize the test accuracy for given communication and computation constraints is an interesting open issue.


\end{document}